\documentclass[referee]{aa}
\usepackage{graphicx}

\def\apj{{Astrophysical Journal}}
\def\apjl{{Astrophysical Journal Letters}}

\def\prd{{Phys. Rev. D}}

\def\pasp{{The Publications of the Astronomical Society of the Pacific}}
\def\mnras{{Monthly Notices of the Royal Astronomical Society}}

\titlerunning{Quark Stars among compact stars}
\authorrunning{Niebergal et al.}

\begin{document}

\title{SGRs and AXPs proposed as ancestors of the Magnificent seven}

\author{Brian Niebergal, Rachid Ouyed, and Denis Leahy }

\institute{Department of Physics and Astronomy, University of Calgary, 
2500 University Drive NW, Calgary, Alberta, T2N 1N4 Canada}

\offprints{bnieber@phas.ucalgary.ca}

\date{recieved/accepted}

\abstract{
The recently suggested correlation between the surface temperature and
 the magnetic field in isolated neutron stars does not seem to work well for
  SGRs, AXPs and X-ray dim isolated neutron stars (XDINs; specifically the
  Magnificent Seven or M7). Instead 
by appealing to a Color-Flavor Locked Quark Star (CFLQS) we 
 find a more natural explanation. In this picture, the heating is provided by
 magnetic flux expulsion from a crust-less superconducting quark star.
  Combined with our previous studies concerning the possibility of SGRs, AXPs, and XDINs as CFLQSs,
   this provides another piece of evidence that these objects are all related.
    Specifically, we propose that XDINs are the descendants of SGRs and AXPs.
\keywords{dense matter --- stars: magnetic fields --- stars: neutron --- 
X-rays: stars --- X-rays: bursts --- radiation mechanisms: non-thermal} }

\maketitle

\section{Introduction}

Heating of neutron stars by magnetic field decay in the crust has been suggested by
Pons et al. (2007) to explain the observed correlation between surface temperature
 and dipolar magnetic field strength of isolated neutron stars. They define a heating balance line (HBL)
 in the temperature-magnetic field diagram.  Since the Soft Gamma-ray Repeaters (SGRs) and 
Anomalous X-ray Pulsars (AXPs) lie well
 above the HBL line and the Magnificent Seven (M7; a group of isolated neutron stars; see for example Haberl 2007) 
fall well below the line, here we explore
  an alternative explanation. Our scenario, rather than crustal field decay,
   involves magnetic flux expulsion from a superconducting crust-less star.
    The most likely compact star that can provide this is a Color-Flavor Locked Quark Star (CFLQS).
More specifically, we employ strange quark matter in the Color-Flavor Locked 
(CFL; Rajagopal \& Wilczek 2001) phase where quarks of certain color and flavor pair together,
resulting in a color superconducting medium.  Due to the rotation of the star,
the medium develops a vortex lattice, where the star's magnetic field is constrained
to reside only inside these vortices (i.e. an Abrikosov lattice).

It is generally accepted
that SGRs and AXPs are the same type of objects, and it has been speculated before that 
X-ray Dim Isolated Neutron stars (XDINs) are also related (see Treves et al. 2000 for a review).
 We have previously proposed (Ouyed et al. 2006b; Ouyed et al. 2006c;
 Niebergal et al. 2006) that Quark Stars in the CFL phase 
not only exist, but are manifested in the form of these three classes of astrophysical objects 
(SGRs, AXPs, and a specific group of XDINs named the M7).
Using our CFLQS model, we present further evidence for the relation between SGRs/AXPs and the M7
based on an analytic prescription for the evolution of the star's effective temperature, $T_{\rm eff}$,  and 
magnetic field strength.  This analytic prescription is derived by considering vortex expulsion
from the star due to spin-down from magnetic braking.
We compare our model with observed
 $T_{\rm eff}$ vs $B$  and find a more natural agreement for SGRs, AXPs, and the M7 
  than the Pons et al. (2007) model.

The paper is presented as follows: In \S~\ref{sec:quarkstars} we
briefly introduce the notion of a CFLQS,
the basis of our model, and the resulting vortex lattice formed within.
We go on to describe the evolution of this vortex lattice in \S~\ref{sec:fluxexpulsion},
and the resulting magnetic flux expulsion due the quantized relation between the total 
number of vortices and the star's spin-period.  Lastly, in \S~\ref{sec:XDINs},
we discuss XDINs (specifically the M7) and give evidence for their ancestral link to SGRs/AXPs.
We then conclude in \S~\ref{sec:conclusion}.

\section{CFL Quark Stars}\label{sec:quarkstars}

We assume a quark star (QS) is born with a temperature $T > T_{\rm c}$
($T_{\rm c}$ is the critical temperature below which superconductivity sets in),
and enters a superconducting-superfluid phase (Color-Flavor Locked; CFL phase) in the core as
it cools by neutrino emission (Ouyed et al. 2002; Ker\"anen et al. 2005), 
and contracts due to spin-down.
The CFL front quickly expands to the entire star followed
by the formation of rotationally induced vortices,
analogous to rotating superfluid $^3$He (the vortex lines
are parallel to the rotation axis; Tilley\&Tilley 1990).
Via the Meissner effect (Meissner \& Ochsenfeld 1933), the magnetic field is partially screened
from the regions outside the vortex cores.
Now the system will consist of, alternating
regions of superconducting material with a screened magnetic
field, and the vortices where most of the magnetic field resides.

As discussed in Ouyed et al. (2004), this has interesting consequences
on how the surface magnetic field adjusts to the interior field which is confined in the vortices.
In Ouyed et al. (2006a) we performed numerical simulations 
of the alignment of a quark star's exterior field, and, found that the physics 
involved was indicative of SGR/AXP activity\footnote{See simulations: 
www.capca.ucalgary.ca/\~{}bniebergal/meissner/}.

\section{Magnetic Dissipation in the Crust vs Flux Expulsion}\label{sec:fluxexpulsion}

Following the initial magnetic field alignment event is the quiescent phase, 
where magnetic braking spins-down the QS, causing the outermost vortices 
to be pushed to the surface and expelled (Ruutu et al. 1997; Srinivasan et al. 1990). 
The magnetic field contained within these vortices is also expelled and
annihilates by means of magnetic reconnection events near the surface of the star,
causing energy release, presumably in the X-ray regime.
The number of vortices decreases slowly as the QS spins down leading to continuous, quiescent,
energy release that can last until the magnetic field is insufficiently strong to produce detectable emission.

\begin{figure*}[t!]
\includegraphics[width=1.\textwidth]{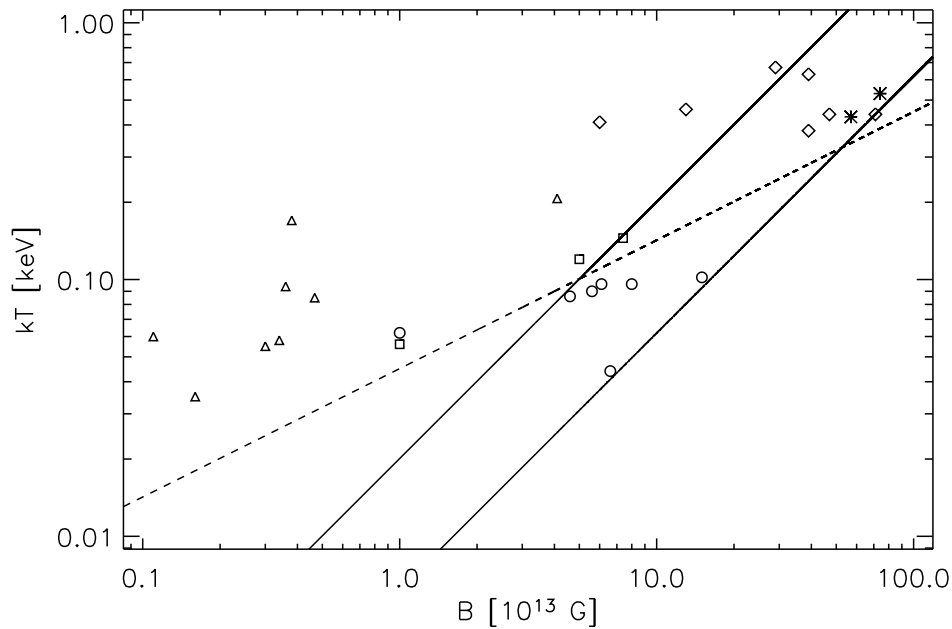}
\caption{\label{fig:temperature}
Effective temperature vs magnetic field (inferred from cyclotron resonance observations where
possible) of isolated neutron stars. 
The stars and diamonds represent SGRs and AXPs respectively,
while radio-quiet, X-ray dim, isolated stars (the magnificent seven)
are shown as circles.  The remaining objects are isolated radio pulsars
with periods $P > 3$ s (squares) and $P < 3$ s (triangles). The dashed line
is the \textit{heating balance line} derived by Pons et al. (2007)
assuming balance between heating by magnetic field decay and blackbody cooling. The
two solid lines are our model of vortex expulsion from a CFL quark star (eq. \ref{eq:eff_temp}). 
The upper and lower lines represent upper
($\Delta R_{\rm km} = 10$, $P_0 = 1.5$ s) and lower
($\Delta R_{\rm km} = 1$, $P_0 = 5$ s) bounds for CFL quark stars.
We point out that the free parameters in our model are tightly constrained by observations.}
\end{figure*}

This scenario differs to what is expected from neutron stars, wherein the proton and neutron superfluids
are thought to compete to push vortices to the surface, where the magnetic field slowly decays as it diffuses through
the neutron star's crust (Konenkov \& Geppert 2000).  Pons et al. (2007) parametrize this field decay and
balance it with blackbody cooling, thus attaining an equilibrium temperature ($T_{\rm eff} \propto B^{1/2}$;
 what Pons et al. referred to as the heat balance line (HBL)) 
below which no neutron stars should be found.  However, most of the Magnificent Seven 
(M7; eg. Haberl 2007) are below this proposed temperature (see Fig.~\ref{fig:temperature}),
or at least require very different physical characteristics than their neutron star counterparts. 

In our CFLQS model, because there is only the CFL matter and no crust\footnote{Although it has been shown that pure CFL matter is rigorously electrically neutral
(Rajagopal \& Wilczek 2001), other work (Usov 2004 and references therein) indicates
that a thin crust is allowed around a quark star due to surface depletion of strange quarks.
In our model we have assumed no depletion of strange quarks, which implies a bare quark star.}, the vortices are efficiently pushed to
the surface where the magnetic field contained within decays by reconnection rather than dissipation.
Thus, by balancing this heating with blackbody cooling we attain an equilibrium temperature
proportional to $B$, rather than $B^{1/2}$.  

This is realized by first considering spin-down due to a rotating, aligned, magnetic dipole, (e.g. M\'esz\'aros 1992,)
\begin{equation}\label{eq:aligned_dipole}
\frac{\dot{\Omega}}{\Omega} \approx -\frac{B^2 R^6 \Omega^2}{I c^3} \ .
\end{equation}
Here, $\Omega$, is the spin frequency, $\dot{\Omega}$, is the spin frequency derivative with respect to time, $B$,
is the magnetic field strength at the surface of the QS, $R$, is the radius of the QS, $I$, is the moment of inertia,
and $c$ is the speed of light.  In the aligned-rotator model the star spins down by
magnetospheric currents escaping through the light cylinder.
 For a neutron star, these currents are thought to originate in the crust.
 Instead, in our model, pair production from magnetic reconnection would likely
  supply the currents (Niebergal et al. 2006). 

From the quantization of angular momentum the number of vortices 
is proportional to the QS's rotation period.
As given in Ouyed et al. (2004), for a sphere this relation is given by
\begin{equation}\label{eq:vortexnum}
\frac{dN_{\rm v}}{d\Omega} \simeq \frac{N_{\rm v}}{\Omega} \ , 
\end{equation}
where $N_{\rm v}$ is the total number of vortices.

Thus, it is easy to imagine that as the star loses rotational energy and spins down,
the QS will lose vortices.
The magnetic field contained within these vortices is also released from the QS.
This implies that the magnetic field possessed by the QS is dependent on the spin period.
However, the rate of spin-down is proportional to the magnetic field squared (cf.~Eq.~\ref{eq:aligned_dipole}),
so the period is also dependent on the magnetic field.
Hence, the spin period and magnetic field are coupled, but they can be solved for independently as 
done by Niebergal et al. (2006), yielding the important relations,
\begin{eqnarray}\label{eq:bp_relation}
\frac{B^2}{\Omega} &=& \frac{B_0^2}{\Omega_0} \nonumber \\ 
P &=& P_0\left(1+t/\tau\right)^{1/3} \nonumber \\ 
B &=& B_0\left(1+t/\tau\right)^{-1/6}  \ ,
\end{eqnarray}
where in the above equations the subscript, $0$, refers to the initial value at the time of the QS's birth.
Also, the characteristic age, $\tau$, in units of years is calculated to be,
\begin{equation}
\tau_{\rm yrs} = 5\times 10^4 \left(\frac{10^{14} {\rm G}}{B_0}\right)^2 
                    \left(\frac{P_0}{5 {\rm s}}\right)^2 
                    \left(\frac{M_{\rm QS}}{M_{\odot}}\right) \left(\frac{10 {\rm km}}{R_{\rm QS}}\right)^4 \ ,
\end{equation}
where, $M_{\rm QS}$, is the mass of the quark star, and, $R_{\rm QS}$, its radius.

As the magnetic field is forced outside of the star, it decays by reconnection, 
causing heating on the surface.  Assuming an efficiency of $10 \%$ for the 
conversion to X-rays from reconnection, then a simple model of heating balanced with cooling
gives an effective equilibrium temperature of the QS to be,
\begin{equation}\label{eq:eff_temp}
  kT_{\rm eff} \simeq 1.4\times 10^{-2} B_{13} \Delta R_{\rm km}^{1/4} P_0^{-1/2} {\rm keV} \ .
\end{equation}
In the above equation, $B_{13}$ is the magnetic field strength at the surface of the QS
in units of $10^{13}$ G, $\Delta R_{\rm km}$ is the size of the emitting region in units
of kilometers, and $P_0$ is the initial period of the QS.

In figure \ref{fig:temperature}, the two solid lines represent the upper 
($\Delta R_{\rm km} = 10$, $P_0 = 1.5$ s) and lower 
($\Delta R_{\rm km} = 1$, $P_0 = 5$ s) bounds for the temperature given by equation (\ref{eq:eff_temp}).
These bounds cover the likely extent of the parameter space, given that $\Delta R_{\rm km}$ 
is unlikely to be larger than the QS itself ($10$ km), and $P_0$ should not be much less than
the SGR/AXP/M7 period average.  Thus, our QS model parameters are very physical, and tightly 
constrained by observations.
The data points are from Pons et al. (2007)
as is the dashed line, which represents their heating balance (HBL) temperature for 
neutron star spin-down combined with a two-parameter best fit for magnetic field 
diffusion through the crust.

We argue that in the context of the HBL model, the M7 would require a very different set of parameters
from other neutron stars; parameters that should be mostly uniform.  
In our QS model, the M7 share a parameter space with SGRs and AXPs, suggesting the two groups
are the same type of objects that differ primarily in age.  The other objects
in figure \ref{fig:temperature} that do not fall within the bounds of our model
are regular neutron stars.  Also, there may in fact be more objects in the gap
between SGRs/AXPs and the M7, but they would likely appear like 
regular X-ray pulsars with no persistent pulsed radio emission,
 of which there may be some unidentified candidates in the ROSAT catalogue.


\section{The XDIN, AXP, \& SGR Link}\label{sec:XDINs}

The M7 are a class of $\sim 10^{6}~\rm{yrs}$ old  
stars possessing relatively strong magnetic field strengths ($10^{13}$ to $10^{14}~\rm{G}$) 
and exhibiting a clustering in their observed periods similar to that of AXPs and SGRs.
 Like AXPs/SGRs they show no persistent pulsed emission in radio wavelengths.
They are also characterized by a near perfect blackbody spectrum (Posselt et al. 2007).
The near perfect blackbody fits naturally within the framework of our model as 
the CFLQS is expected to possess no crust, but rather a bare surface.  Moreover,
the lack of radio pulsations, in our model, is a necessary consequence 
of the birth of a CFLQS, which causes the star's magnetic field to align with its rotation axis (Ouyed
et al. 2006a).

Although XDINs have previously been speculated to be related to AXPs and SGRs 
(see Treves et al. 2000 for a review), the most popular neutron star models were 
unable to explain period clustering and sustainment of the magnetic field.
In our CFLQS model, after the QS's field has aligned, 
it will spin-down through magnetic braking as described in equations (\ref{eq:bp_relation}),
and for ages on the order of $\sim 10^{6}~\rm{yrs}$, we arrive at results indicative of the M7. 
As an example, if a CFLQS is born with a radius of $9$ km,
period of $P_0 = 3~\rm{s}$, and magnetic field strength of $B_0 = 10^{14}~\rm{G}$, 
then by the time it reaches ages estimated for XDINs it will
have attained a period of $10~\rm{s}$ and its field will
have decayed to $\sim 5\times 10^{13}~\rm{G}$ (see Fig. 3
 in Niebergal et al. 2006).  

Hence, by using our model with SGR/AXP parameters we arrive at M7 parameters
after roughly less than a million years, suggesting age is the primary difference 
between SGRs/AXPs and the M7. Also, after roughly $5\times 10^4$ years the 
magnetic field (and resulting luminosity) begins to drops
rapidly in our model, implying no XDIN beyond that age should be detectable,
unless it is very close in distance (Niebergal et al. 2006).
Recent estimates of distances to the M7 (Posselt et al. 2007) satisfy this criteria,
as the distance varies from roughly 160 to 400 parsecs.
These new distance estimates also seem to indicate that the number of currently observable
SGRs/AXPs is consistent with the seven observed\footnote{Despite intensive searches for more objects similar to the M7, none have been
found since 2001 (Haberl 2007).} XDINs, given their ages.

It is worthwhile to point out that, in our model, the dipole magnetic field strength is given by 
$B_{\rm d} = 3\times10^{19} \sqrt{3 P\dot{P}}$ G, which is greater than the usual
field estimation for neutron stars by a factor of $\sqrt{3}$.
  Thus, vortex expulsion changes the braking index of a spinning dipole ($n \rightarrow 4$),
and results in an extra factor of $\sqrt{3}$ when predicting the star's magnetic field strength
from its spin-period and spin-down rate.
This extra factor may help account for the discrepancy between the M7's estimated dipole field 
(using the usual neutron star braking index; $n=3$) and the observed field using 
cyclotron resonances (eg. Haberl 2007). 

\section{Conclusion}\label{sec:conclusion}

We have shown in this Letter that, in the context of CFLQS model 
many features of the Magnificent Seven  can be explained and evidence for their 
SGR/AXP ancestry was presented.  Specifically the evolution of:  i) The effective temperature; 
ii) spin-period; and iii) the magnetic field are all predicted for 
SGRs/AXPs and the M7 using our CFLQS model.
A CFLQS also has the advantage of not possessing a crust, thus its bare surface is able
to explain the near featureless spectrum of the M7 (Pons et al. 2005).  The SGR/AXP spectrum is naturally more
convoluted as they are much further away, and so would likely suffer from interstellar absorption effects.
 Other properties of XDINs such as; (i) the two-component blackbody,
(ii) the optical excess, and (iii) the absorption lines have been discussed in Ouyed et al. (2006b)
 in the context of CFLQS (see also Ouyed et al. 2006c).  While we have suggested
  possible evolutionary signatures of quark stars among neutron stars, signatures
  of their birth might have already been seen (Leahy\&Ouyed 2007).


\begin{acknowledgements}
This research is supported by grants from the Natural Science and
Engineering Research Council of Canada (NSERC).
\end{acknowledgements}


\clearpage

\end{document}